\documentclass[12pt,twoside]{article}
\usepackage[latin1]{inputenc}
\usepackage{graphicx}
\usepackage{a4wide}
\usepackage{listings}
\usepackage{amsmath}

\newcommand{\ee}{$e^{+}e^{-}$}
\newcommand{\NN}{$\textrm{N}$+$\textrm{N}$}
\newcommand{\pnb}{$\textrm{p}$+$\textrm{Nb}$}

\newcommand{\cc}{$\textrm{C}$+$\textrm{C}$}
\newcommand{\caca}{$\textrm{Ca}$+$\textrm{Ca}$}
\newcommand{\arkcl}{$\textrm{Ar}$+$\textrm{KCl}$}
\newcommand{\np}{$\textrm{n}$+$\textrm{p}$}
\newcommand{\pp}{$\textrm{p}$+$\textrm{p}$}{\LARGE }

\newcommand{\xdp}{$\textrm{d}$+$\textrm{p}$}
\newcommand{\gevu}{G$e$V/$u$}
\newcommand{\gevcc}{G$e$V/$c^{2}$}

\newcommand{\gev}{G$e$V}
\newcommand{\mev}{M$e$V}
\newcommand{\etal}{$et~al.$}

\begin{document}

\begin{center}

\section*{Probing resonance matter with virtual photons\footnote{Talk given at International Nuclear Physics Conference - INPC 2010, July 4 - 9 2010, University of British Columbia, Vancouver, Canada}}

\thispagestyle{empty}

\markboth{Tetyana Galatyuk {\it et al.}}
{Probing resonance matter with virtual photons}

T.~Galatyuk$^{7}$, G. Agakishiev$^{6}$, A.~Balanda$^{3}$, D.~Belver$^{15}$, A.V.~Belyaev$^{6}$, A.~Blanco$^{2}$,
M.~B\"{o}hmer$^{11}$, J.~L.~Boyard$^{13}$, P.~Braun-Munzinger$^{4,b}$, P.~Cabanelas$^{15}$
% ...

\vspace{0.35cm}
(HADES collaboration)
\vspace{0.35cm}

\renewcommand{\baselinestretch}{0.9}
\footnotesize
\textit{
\hspace{-0.4cm}\makebox[0.3cm][r]{$^{1}$}
Istituto Nazionale di Fisica Nucleare - Laboratori Nazionali del Sud, 95125~Catania, Italy\\
\hspace{-0.4cm}\makebox[0.3cm][r]{$^{2}$}
LIP-Laborat\'{o}rio de Instrumenta\c{c}\~{a}o e F\'{\i}sica Experimental de Part\'{\i}culas, 3004-516~Coimbra, Portugal\\
\hspace{-0.4cm}\makebox[0.3cm][r]{$^{3}$}
Smoluchowski Institute of Physics, Jagiellonian University of Cracow, 30-059~Krak\'{o}w, Poland\\
\hspace{-0.4cm}\makebox[0.3cm][r]{$^{4}$}
GSI Helmholtzzentrum f\"{u}r Schwerionenforschung GmbH, 64291~Darmstadt, Germany\\
\hspace{-0.4cm}\makebox[0.3cm][r]{$^{5}$}
Institut f\"{u}r Strahlenphysik, Forschungszentrum Dresden-Rossendorf, 01314~Dresden, Germany\\
\hspace{-0.4cm}\makebox[0.3cm][r]{$^{6}$}
Joint Institute of Nuclear Research, 141980~Dubna, Russia\\
\hspace{-0.4cm}\makebox[0.3cm][r]{$^{7}$}
Institut f\"{u}r Kernphysik, Goethe-Universit\"{a}t, 60438 ~Frankfurt, Germany\\
\hspace{-0.4cm}\makebox[0.3cm][r]{$^{8}$}
II.Physikalisches Institut, Justus Liebig Universit\"{a}t Giessen, 35392~Giessen, Germany\\
\hspace{-0.4cm}\makebox[0.3cm][r]{$^{9}$}
Istituto Nazionale di Fisica Nucleare, Sezione di Milano, 20133~Milano, Italy\\
\hspace{-0.4cm}\makebox[0.3cm][r]{$^{10}$}
Institute for Nuclear Research, Russian Academy of Science, 117312~Moscow, Russia\\
\hspace{-0.4cm}\makebox[0.3cm][r]{$^{11}$}
Physik Department E12, Technische Universit\"{a}t M\"{u}nchen, 85748~M\"{u}nchen, Germany\\
\hspace{-0.4cm}\makebox[0.3cm][r]{$^{12}$}
Department of Physics, University of Cyprus, 1678~Nicosia, Cyprus\\
\hspace{-0.4cm}\makebox[0.3cm][r]{$^{13}$}
Institut de Physique Nucl\'{e}aire d'Orsay, CNRS/IN2P3, 91406~Orsay Cedex, France\\
\hspace{-0.4cm}\makebox[0.3cm][r]{$^{14}$}
Nuclear Physics Institute, Academy of Sciences of Czech Republic, 25068~Rez, Czech Republic\\
\hspace{-0.4cm}\makebox[0.3cm][r]{$^{15}$}
Departamento de F\'{\i}sica de Part\'{\i}culas, University of Santiago de Compostela, 15782~Santiago de Compostela, Spain\\
\hspace{-0.4cm}\makebox[0.3cm][r]{$^{16}$}
Instituto de F\'{\i}sica Corpuscular, Universidad de Valencia-CSIC, 46971~Valencia, Spain\\
\hspace{-0.4cm}\makebox[0.3cm][r]{$^{a}$}
{also at ISEC Coimbra, ~Coimbra, Portugal}\\
\hspace{-0.4cm}\makebox[0.3cm][r]{$^{b}$}
{also at ExtreMe Matter Institute EMMI, 64291~Darmstadt, Germany}\\
\hspace{-0.4cm}\makebox[0.3cm][r]{$^{c}$}
{also at Technische Universit\"{a}t Dresden, 01062~Dresden, Germany}\\
\hspace{-0.4cm}\makebox[0.3cm][r]{$^{d}$}
{also at Dipartimento di Fisica, Universit\`{a} di Milano, 20133~Milano, Italy}\\
\hspace{-0.4cm}\makebox[0.3cm][r]{$^{e}$}
{also at Panstwowa Wyzsza Szkola Zawodowa , 33-300~Nowy Sacz, Poland}
}
\renewcommand{\baselinestretch}{1}
\normalsize

\newpage
\textbf{Abstract}
\vspace{0.5cm}\\
\begin{minipage}[c]{0.9\columnwidth}
In the energy domain of $1-2$ \gev\ per nucleon, HADES has measured rare penetrating probes (\ee) in \cc, \arkcl, \pp, \xdp\ and \pnb\ collisions. For the first time the electron pairs were reconstructed from quasi-free \np\ sub-reactions by detecting the proton spectator from the deuteron breakup. An experimentally constrained \NN\ reference spectrum was established. Our results demonstrate that the gross features of di-electron spectra in \cc\ collisions can be explained as a superposition of independent \NN\ collisions.  On the other hand, a direct comparison of the \NN\ reference spectrum with the \ee\ invariant mass distribution measured in the heavier system \arkcl\ at $1.76$ \gevu\ shows an excess yield above the reference, which we attribute to radiation from resonance matter. Moreover, the combined measurement of di-electrons and strangeness in \arkcl\ collisions has provided further intriguing results which are also discussed.
\end{minipage}
\end{center}

\section{Phases of dense quark-gluon matter}
\begin{figure}[tb]
\begin{center}
\includegraphics[width=0.87\textwidth]{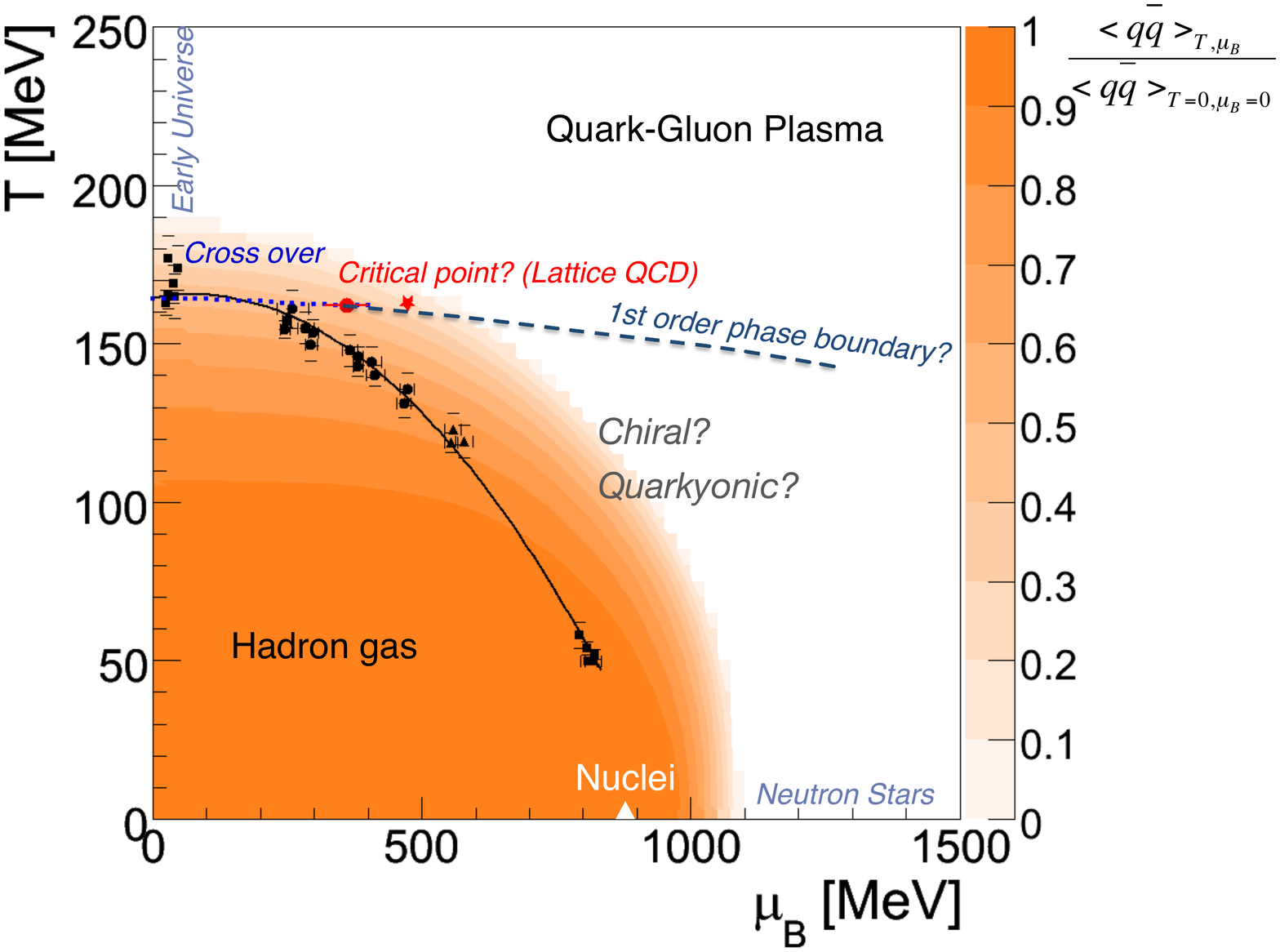}
\caption{The phase diagram of QCD including data points in $T$ and $\mu_{B}$ describing the final hadron abundancies in a statistical model~\cite{cleymans_qcd}. The normalized chiral condensate $\frac{<q\bar{q}>_{T,\mu_{B}}}{<q\bar{q}>_{T=0,\mu_{B=0}}}$  in dependence on $T$ and $\mu_{B}$ is shown as $3^{rd}$ dimension in color code~\cite{wambach}. The condensate ratio is reduced for both high $T$ and $\mu_{B}$ .}
\label{qcd_phase}
\end{center}
\end{figure}

During the recent years our knowledge about strongly interacting matter greatly advanced. Fundamental aspects of matter, from its creation in the early universe to phases of matter in compact stellar objects are addressed.  High-energy heavy-ion collisions provide a tool to study strongly interacting matter under extreme conditions, i. e.\ densities and temperatures.  A qualitative picture of the phase diagram of nuclear matter as a function of temperature ($T$) and baryochemical potential ($\mu_{B}$) is shown in Fig.~\ref{qcd_phase}.

Current and future experiments at RHIC at the BNL and at LHC at CERN concentrate on the study of matter at very high temperatures and at low baryochemical potential, i.e.\ in the region where a smooth crossover from a deconfined to a hadronic phase is predicted by lattice QCD calculations. The accelerator facilities AGS at BNL and the SPS at CERN accessed extreme states of matter, as well, at high temperature but at higher net-baryon densities. At even lower bombarding energies (E$_{lab}\simeq1-2$~\gevu) at the Bevalac and the current GSI SIS18 facilities, a large region in the nuclear matter phase diagram ranging from ground state matter density $\rho_{0}$ up to about $3\rho_{0}$ can be accessed with a proper choice of the collision system. There, the reaction volume is heated up to temperatures $T\leq80$~\mev, very likely without reaching the QGP phase boundary. On the other hand, the Nambu Jona-Lasinio model predicts a substantial depletion of the chiral condensate already at SIS18 energies. Such an exotic state of matter is distinct from the confined and deconfined phases and represents the matter which is confined, yet chirally symmetric, as has been suggested in~\cite{larry_mc_lerran1,andronic} as a new phase of QCD, called Quarkyonic matter.  
\newpage
\section{Probes of extreme matter}
The challenge remains to detect these phases in the laboratory by isolating unambiguous signals. Among the promising observables for investigations are short-lived vector mesons decaying into lepton pairs inside the hot and dense matter. Such purely leptonic final states carry important information on the decaying objects to the detectors without being affected by strong final-state interaction while traversing the medium. It also turned out that the sensitivity of the strange particle probe to medium properties is strongly enhanced if the beam energy is below the strange particle production threshold energy in nucleon-nucleon collisions. This is the case in nucleus-nucleus collisions in the SIS energy regime for some reaction channels.

At beam energies of $1 - 2$~\gevu\ di-electron production was studied by the DLS\footnote{DiLepton Spectrometer} collaboration at the Bevalac~\cite{dls_prl_porter}. A large di-electron excess over the "hadronic cocktail" was observed in \cc\ and \caca\ collisions. However, in contrast to the high-energy experiments~\cite{ceres_pb_au_158_7per_centrality,na60_prl}, for a long time the excess could not be explained by any theoretical model and the situation hence became famous as the $"$DLS puzzle$"$. Medium effects at moderate energies are closely linked to the effects at high beam energies through VMD\footnote{Vector Meson Dominance}. At SIS/Bevalac energies the pions hide themselves by forming excited baryon states ($\Delta$,~N$^*$) which carry the produced heat to be released in subsequent collisions. Finally, at a late stage, the pion-depleted system releases the pions via decay of the resonances. It has been established that baryon-driven medium effects are the key in describing the low-mass SPS data (for a review see~\cite{rapp}). The broadening mass scenario implies a strong coupling of the $\rho$ meson to baryons. Understanding these contributions is important for low-mass lepton pair program at FAIR (also at RHIC and even at LHC). Besides the resonance decays, a strong bremsstrahlung contribution in \np\ interactions has been predicted within the framework of a covariant OBE\footnote{One Boson Exchange} model~\cite{kaptari}. Finally, the question whether the observed excess of dileptons is related to any in-medium effect remains open because of uncertainties in the description of elementary di-electron sources. Precise understanding of the elementary reactions is therefore important for the interpretation of di-electron emission in heavy-ion collisions.

At incident energies around $1-2$ \gevu, particles carrying strangeness have turned out to be very valuable messengers too. Over the last $15$ years, the KaoS and FOPI experiments at GSI performed measurements on different aspects of the strange particle physics. KaoS data indicate that the yield of K$^{+}$ mesons produced in heavy-ion collisions is affected by both the nuclear compressibility and the in-medium kaon potential~\cite{kaos}. In nucleon-nucleon collisions the K$^{+}$ multiplicity exceeds by more than an order of magnitude the K$^{-}$ multiplicity at the same Q~value. In nucleus-nucleus collisions, however, production of kaons and antikaons turns out to be equal. Such a behavior is explained in transport models  by the effect of strangeness exchange reactions~\cite{kaons_bratkovskaya}. Results, recently obtained by FOPI and HADES for the $\phi$/K$^{-}$ ratio, will be used to check this conclusion.
\section{The HADES strategy}
The major experimental challenge is to discriminate the penetrating but very rare leptons from the huge hadronic background which exceeds the electron signal by many orders of magnitude. The HADES\footnote{High Acceptance Di-Electron Spectrometer}~\cite{hades_tech} detector has been specifically designed to overcome these difficulties. It is set up at SIS18 at the GSI Helmholtzzentrum f\"ur Schwerionenforschung (Germany) to study \ee\ and hadron production in heavy-ion, as well as elementary and pion-induced reactions in the energy regime of $1-2$~\gevu\ in a systematic way, with high quality data.
\subsection{Intermediate mass excess in \cc\ and  \arkcl\ collisions}
The excess of electron pairs in \cc\ collisions was investigated by the HADES experiment for beam energies of $1$ and $2$~\gevu~\cite{hades_cc1gev,hades_cc2gev}. The excitation function of the excess-pair multiplicity with masses larger than those from $\pi^0$ Dalitz decays, i.e.\ in the mass range from $0.15$~\gevcc\ to $0.5$~\gevcc, scales like the $\pi^{0}$ multiplicity. This fact provides a hint to the possible origin of the excess yield; pion production at these low energies is known to be dominantly coming from the excitation and decay of baryonic resonances (mainly the $\Delta(1232)$ resonance). To contribute to a better understanding of the contributions to di-electrons from the early stage of heavy-ion collisions, HADES has studied \pp\ and \xdp\ interactions at E$_{kin} = 1.25$ \gevu, i.e.\ below the $\eta$ meson production threshold in proton-proton reactions~\cite{hades_nnckt}. The main goal of the latter experiment was to understand the \np\ bremsstrahlung component for \ee\ production in the tagged reaction channel $np \rightarrow npe^{+}e^{-}$ and to establish an experimental cocktail of di-electrons from $"$free$"$, i.e.\ $"$non-medium$"$ hadron decays at SIS energies.
\begin{figure}[tb]
\begin{center}
\includegraphics[width=0.87\textwidth]{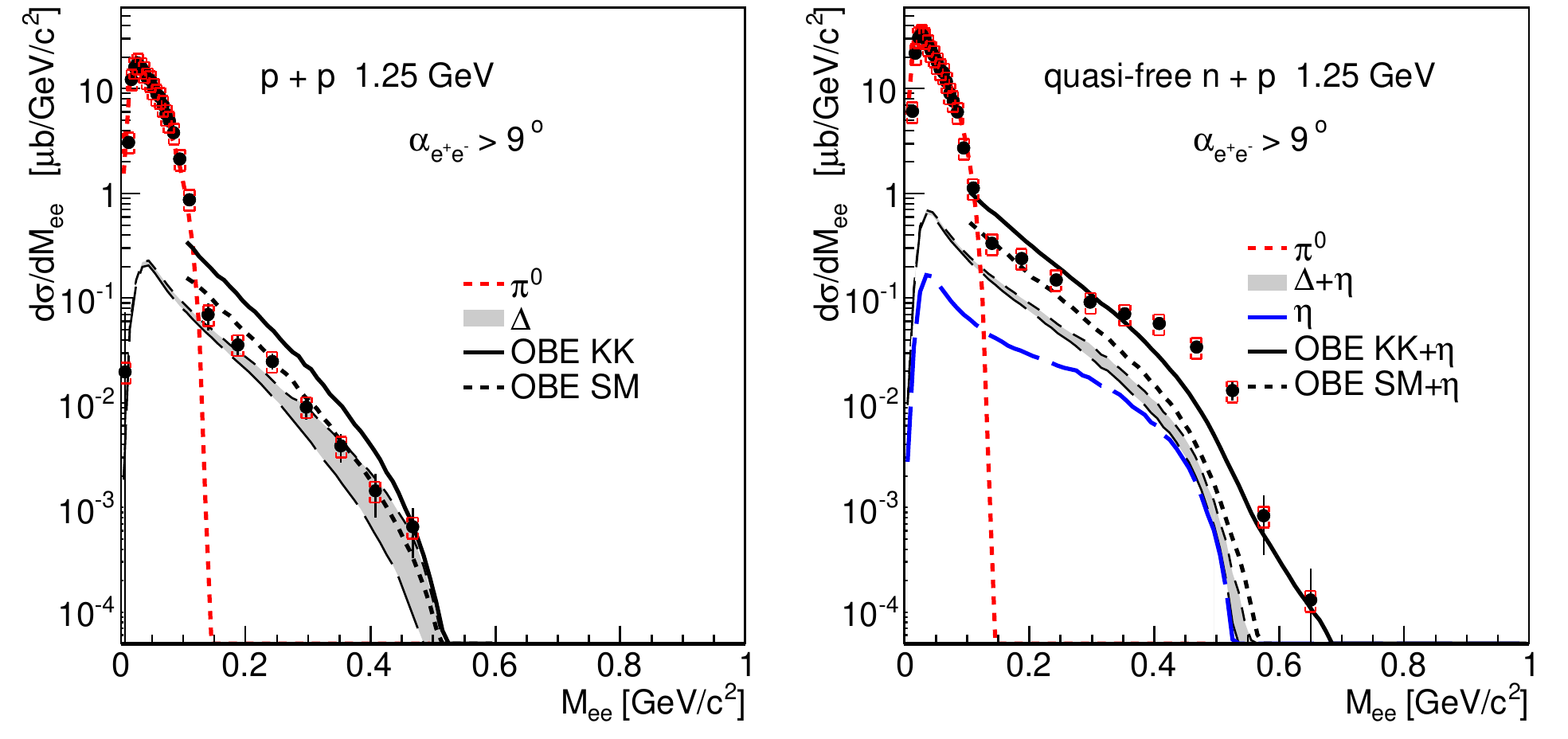}
\caption{Left: Electron pair differential cross sections as function of invariant mass (full circles) measured in \pp\ reactions. Right: Same in quasi-free \np\ reactions at $1.25$ \gev. Systematic errors (constant in the whole mass range) are indicated by (red) horizontal bars, statistical errors by vertical bars. In the analysis, \ee\ pairs with an opening angle of $\alpha > 9^\circ$ were removed from the sample. The lines show results of model calculations with the Pluto event generator (for a review see~\cite{pluto}).}
\label{pp_np}
\end{center}
\end{figure}
The inclusive cross section for electron pair production in \pp\ and quasi-free \np\ collisions at $1.25$ \gev\ as a function of the pair invariant mass is shown in Fig.~\ref{pp_np}. Our data demonstrate a very strong isospin dependence of the di-electron production. In the reaction $np \rightarrow  np e^+e^-$, extracted from the tagged subreaction in $dp \rightarrow p_{spectator}np e^+e^-$, at intermediate values of the di-electron invariant mass a shoulder is visible. It is not a trivial task to interpret such a structure.  OBE model calculations~\cite{kaptari,shyam} show that the bremsstrahlung and the $\Delta$ Dalitz contributions appear to be almost equally important for \np\ collisions, while for \pp\ collisions the $\Delta$ decay plays the dominant role and bremsstrahlung is strongly suppressed. The OBE model of Ref.~\cite{shyam} reproduces the \pp\ experimental data in the whole mass range quite well, while still using certain outdated parameterizations of the bremsstrahlung. The OBE model of Ref.~\cite{kaptari} can reproduce the shape of the \pp\ mass spectra as well, however, has a problem with the overall normalization. Comparison of the \np\ data with OBE models (see Fig.~\ref{pp_np}, right panel) shows that both models fail in reproducing the quasi-free \np\ data in the mass region above the $\pi^{0}$ Dalitz decay range. It is important to mention that subtle interference effects combined with a resonant behavior of the transition form factor due to VMD could introduce a strong energy dependence. With those effects the tuned OBE calculations~\cite{shyam2010} can reproduce \np\ data reasonably well. However, in the available transport model calculations of dilepton production in \NN\ collisions interference effects are not included yet. Those calculations fail indeed in reproducing the quasi-free \np\ data~\cite{bratkovskaya,galatyuk_cpod09}. Understanding the elementary channels therefore remains challenging.

\begin{figure}[tb]
\begin{center}
\includegraphics[width=0.87\textwidth]{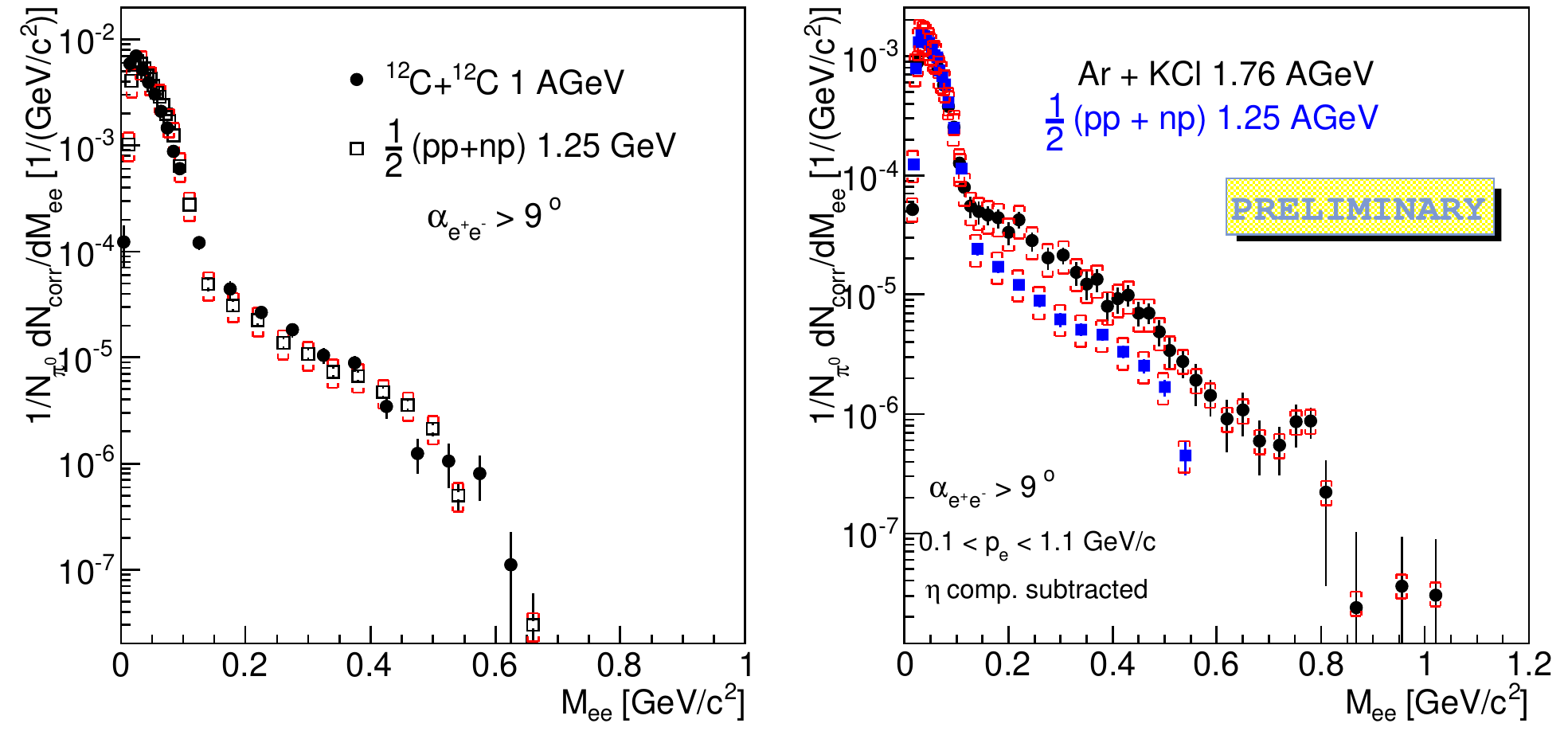}
\caption{Left: Normalized electron pair yield as a function of invariant mass measured in \cc\ collisions (full circles) at 1 \gevu\ compared to the reference yield obtained from \pp\ and \np\ collisions (open squares). Statistical and systematic errors of the measurement are shown as vertical and horizontal bars, respectively. Right: Comparison of the reference spectrum from \NN\ collisions with the reconstructed \ee\ mass distribution in \arkcl. Contributions from the $\eta$ Dalitz decay have been subtracted here.}
\label{nn_cc_arkcl}
\end{center}
\end{figure}

In view of this strong isospin dependence of the dilepton yields in \NN\ collisions the questions arises, whether the \cc\ data can be explained by a superposition of individual \NN\ collisions. Normalizing the efficiency corrected $\frac{1}{2}(pp+pn)$ spectrum to $\pi^0$ and comparing it to the \cc\ invariant-mass spectrum measured at $1$~\gevu\ beam energy shows that the superposition of elementary collisions is sufficient to describe the gross features of the observed di-electron yield in \cc\ collisions (see Fig.~\ref{nn_cc_arkcl}, left panel). This means that the unknown pair source, which for more then ten years has been at the origin of the so-called $"$DLS puzzle$"$, is already present in the elementary \NN\ reactions and is of baryonic origin. 

It has been shown that after normalizing the spectra to the $\pi^{0}$ yield and subtracting the $\eta$ Dalitz contribution (taken from~\cite{taps_eta}), the  \NN\ reference spectrum compares well with the \cc\ data even at slightly defferent bombarding energies~\cite{hades_nnckt}. Moreover, the $"$pion-like$"$ beam-energy dependence of the excess suggests that it is related to baryonic resonance dynamics. This is not longer true when going to a larger system, e. g.\ the \arkcl. After the two previous heavy-ion \cc\ runs~\cite{hades_cc1gev,hades_cc2gev}, \arkcl\ is the heaviest nucleus-nucleus system which has been studied so far with the HADES spectrometer. In the \arkcl\ run we collected 2 billion of first-level-trigger events. The mass resolution, at the $\omega$ pole mass, reaches less than 3$\%$. The di-electron invariant mass distribution shows for the first time a clear $\omega$ signal at SIS energies~\cite{filip_qm09}. A direct comparison of the \NN\ reference spectrum with the invariant-mass distribution measured in \arkcl\ collisions at 1.76~\gevu\ shows a pronounced yield enhancement in the $"$excess region$"$, pointing to a non-linear dependence on the number of participants. This is demonstrated in Fig.~\ref{nn_cc_arkcl} (right panel).

Yet another, experimentally unresolved question, is related to the cold nuclear matter effects~\cite{kek,jlab,taps}. Our experiments were motivated by the ground breaking studies on the role of spontaneous symmetry breaking in hadronic mass generation introduced by R.~D.~Pisarski (1982). Later, the investigation of the vector meson properties in cold medium has focussed on effects due to a direct coupling of the vector meson to resonance-hole states excitations~\cite{friman}. To contribute to this problem, the $\omega$ meson has been produced and identified in dedicated experiments, investigating the \pp\ and  \pnb\ reactions at $3.5$~\gev. The measured lepton pair distribution shows a clear $\omega$ peak with a mass resolution of about 2$\%$ and an signal to background ratio S/B $\leq 10$~\cite{weber_meson10}. This spectrum will serve as a reference for further (in-medium) studies of the line shape in \pnb\ reactions.
\subsection{Strange hadrons measured at sub-threshold energy}
Strange hadrons, i. e.\  K$^{-}$, K$^{+}$, K$^{0} \rightarrow \pi^{+} \pi^{-}$, $\Lambda \rightarrow p \pi^-$, $\phi \rightarrow \textrm{K}^{+}\textrm{K}^{-}$, $\Xi^{-} \rightarrow \Lambda \pi^{-}$, have also been reconstructed in \arkcl\ collisions. The measured yields and slopes of the transverse mass of kaons and $\Lambda$ have been found in agreement with former results of the KaoS and FOPI experiments obtained in similar collision systems. The production of $\phi$ meson and double-strange $\Xi^{-}$ hyperon, however, shows unexpected features. Please note that the $\Xi(1321)$ hyperon was measured 640 \mev\ below its free \NN\ threshold. The statistical hadronization model does not describe the measured $\Xi^{-}$ yield~\cite{hades_xi}. At the same time, it provides a reasonably good estimation of the  $\phi$ meson multiplicity without strangeness suppression. A large $\phi / \textrm{K}^{-}$ ratio of 0.37$\pm$0.13, observed by both FOPI~\cite{fopi_phik} and HADES ~\cite{hades_phik,lorenz}, indicates the importance of the $\phi$ meson in subthreshold K$^{-}$ production. For a firm understanding of the strangeness dynamics the feeding of K$^{-}$ from $\phi$ decays needs to considered~\cite{kaempfer_schade} which supplements strangeness exchange reactions as the dominant kaon production mechanism.
\section{R\'{e}sum\'{e} and prospects}
HADES provides high-quality data for an understanding of the di-electron and strangeness production in elementary and heavy-ion collisions at SIS energy regime. The previously puzzling di-electron excess in the intermediate mass range of observed in \cc\ collisions at  $1$ and $2$ \gevu\ can be described by a superposition of elementary \pp\ and \np\ collisions. The excess has been traced back essentially to effects present already in \np\ collisions.

The \arkcl\ experiment provided data for a moderately large collision system, allowing to continue our studies of dilepton production started with the light \cc\ system. We can now address the evolution not only with beam energy, but also with system size of the so-called pair excess, first observed by the DLS in the mass range of $0.15 - 0.50$ \gevcc. Dilepton spectra measured in \arkcl\ collisions show a significant in-medium radiation in the $"$excess region$"$. Moreover, for the first time at SIS energies, a clear $\omega \rightarrow e^{+}e^{-}$ signal was observed. Analyses of the experimentally obtained hadronic yields measured in \arkcl\ collisions at a E$_{beam} = 1.76$ \gevu\ have shown that the data can be well reproduced within statistical model (except in the case of the $\Xi^{-}$) allows to fix chemical freeze-out coordinates in $T - \mu_{B}$ phase diagram of strongly interacting matter.

Further investigations to search for significant medium effects, based on the established \NN\ reference spectrum, are planned by HADES and will concentrate on heavier collision systems, i.e. $\textrm{Ag}+\textrm{Ag}$ and $\textrm{Au}+\textrm{Au}$. The ultimate goal is to provide a complete excitation function of dilepton production up to energies of $8$~\gevu: ranging from SIS18 (covering energies up to $2$~\gevu) and later on to SIS100 taking over for the higher beam energies at FAIR.
\section*{Acknowledgments}
\renewcommand{\baselinestretch}{0.9}
\footnotesize
The collaboration gratefully acknowledges the support  by Helmholtz Alliance HA216/EMMI, by BMBF grants 06TM970I, 06GI146I, 06F-140, and 06DR135 (Germany), by GSI (TM-FR1,GI/ME3,OF/STR), by grants GA CR 202/00/1668 and GA AS CR IAA1048304 (Czech Republic), by grant KBN 1P03B 056 29 (Poland), by INFN (Italy), by CNRS/IN2P3 (France), by grants MCYT FPA2000-2041-C02-02 and XUGA PGID T02PXIC20605PN (Spain), by grant UCY-10.3.11.12 (Cyprus), by INTAS grant 03-51-3208 and by EU contract RII3-CT-2004-506078.
\renewcommand{\baselinestretch}{1}
\normalsize


\end{document}